\documentclass[journal]{IEEEtran}
\usepackage{cite}
\usepackage[pdftex]{graphicx}
\usepackage[cmex10]{amsmath}
\usepackage{amssymb}
\usepackage{array}
\usepackage{fixltx2e}
\usepackage{color}
\usepackage{url}

\begin{document}

\title{Understanding Biometric Entropy and Iris Capacity:
    Avoiding Identity Collisions on National Scales}
\author{John Daugman$^{1}$ 
 \thanks{$^{1}$Dept.\ of Computer Science and Technology, Cambridge University, UK.
 www.CL.cam.ac.uk/users/jgd1000/  \ \ \  E-mail:  John.Daugman@CL.cam.ac.uk} 
}
\markboth{  }
{Biometric Entropy and Iris Capacity} 

\maketitle
\begin{abstract}
The numbers of persons who can be enrolled by their iris patterns
with no identity collisions is studied in relation to the biometric
entropy extracted, and the decision operating threshold.  The population
size at which identity collision becomes likelier than not, given those
variables, defines iris ``capacity."  The general solution to this combinatorial
problem is derived, in analogy with the well-known ``birthday problem."
Its application to unique biometric identification on national 
population scales is shown, referencing empirical data
from US NIST (National Institute of Standards and Technology) trials involving
1.2~trillion ($1.2 \times 10^{12}$) iris comparisons.  The entropy of a
given person's two iris patterns suffices for global identity uniqueness.
\end{abstract}

\section{Introduction}

\IEEEPARstart{A}{pplicants} for Cambridge University undergraduate studies 
in mathematics or computer science are asked sometimes in their College
interviews to reason about the ``birthday problem":  how many people,
chosen at random, must be assembled until it becomes more likely than not
that at least one pair of them have the same birthday?  Some students
are surprised that the answer is only 23 people.  Although arriving
at the exact number requires a calculator, the reasoning is that 
$N$ people make $N(N-1)/2$ possible pairings.  Given that each pairing
has probability $1/365$ of sharing their birthday and $364/365$ of not,
the probability that {\em none} of the pairings share a birthday is
approximately $(364/365)^{N(N-1)/2}$, which is $< 0.5$ once $N \ge 23$.

There is a clear analogy with biometric collision avoidance, which we
can formulate as the: 
\begin{quote}
{\bf Biometric birthday problem}: if some biometric technology is operating
with a verification FMR (``one-to-one" False Match Rate), how many 
people, chosen at random, must be assembled until it becomes more likely
than not that at least one pair of them have a biometric collision
(are falsely matched to each other)? 
\end{quote}

A good example is face recognition, tested across a broad variety of
scenarios and using a wide range of image quality, for which a good 
performance benchmark corresponds to making just one verification False Match
in 1,000 non-mated comparisons \cite{Ira}\cite{Grother-face}\cite{PNAS2018}. 
That accuracy standard is better than human (even ``super-recogniser")
performance in some circumstances \cite{PNAS2018}.  Face recognition
algorithms have improved greatly in recent years, in terms of Rank-1
identification rates \cite{Ira}\cite{Grother-face} in test protocols in which
a correct match does always exist within a search gallery that is populated
also with other ``distractors".  But even in the recent tests, the best
algorithms do still make some False Matches to distractor images even when
there are only 100 distractors \cite{Ira}\cite{Grother-face} despite the
presence of a correct match within the gallery, that should instead actually
be returned at Rank-1.

Let us now consider the ``biometric birthday problem" for a face recognition
algorithm performing at FMR = 0.001 when examining a gallery of non-mated
faces.  How large must this gallery get before False Matches become
likelier than not, in all-versus-all comparisons?  The answer:  just 38.
That number creates $38 \cdot 37/2 = 703$ possible pairings to consider,
and $(1 - 0.001)^{703} = 0.495$ so False Matches are then already likelier
than not.  When waiting at Passport Control (or some other such queue),
it is entertaining to turn around, look at the first 38 persons standing
behind oneself, and try to spot the pair of facial doppelg\"{a}ngers
\cite{doppels} among them.

Biometric deployments at a national or even prospectively at the planetary
scale face a massively challenging biometric ``birthday problem" if they need
to search for any duplicate identities, as was necessary in India when all
1.4~billion citizens were recently enrolled in a national ID programme
for welfare distribution, government services, and subsidies (UIDAI:  
Unique IDentification Authority of India) \cite{Aiyar}. Because
enrollees had an incentive to acquire multiple identities and thereby
issuance of multiple subsidies, every new enrollment had to be compared
against all existing enrollments before an Aadhaar would be issued.
This amounts to a search for identity collisions, all-versus-all, among
an astronomical $N(N-1)/2$ pairings of persons.  Obviously any attempt 
to do this by face recognition would drown in False Matches from the
very beginning.  There simply is not enough entropy, or randomness, in
human face structure; the necessary functional purposes of major facial
features (mouth, nose, ocular areas) constrain their possible randomness.
The bilateral symmetry normally present in a face further reduces its
entropy by half.  The key idea, the fundamental factor underlying the
power of biometric identification, is entropy \cite{CovThom} \cite{Daug2015}.

Weak biometrics may be sufficient to enable ``one-to-one" verification;
stronger biometrics may enable identification in a search database
of size $N$, ``one-to-few" or ``one-to-many" depending on $N$;
but de-duplication applications exemplify the birthday problem in that
they are essentially ``all-versus-all", and the number of False Match
opportunities they must survive grows massively with $N$.  In such
deployments on a national scale, falsely detected or undetected identity
collisions (even if few in percentage) would lead to reduced public
confidence in and acceptance of the system, its impaired functionality,
and legal problems caused both by undetected duplicates and falsely
detected ones.  Table~I presents, for a broad range of FMR levels
spanning 15 orders-of-magnitude, how large $N$ can get before collisions
become likelier than not.   Table~I clearly shows that the demands for
a minuscule FMR become extremely daunting once the population
size $N$ is even that of a small town, let alone a population of
national, continental, or of planetary scale.

\begin{table}
\renewcommand{\arraystretch}{1.3}
\caption{Accuracy Requirements for Biometric Collision Avoidance}
\centering
\begin{tabular}{|c|r|}    \hline
 Verification FMR  & Critical Population Size $N$ \\ \hline \hline
0.001 & 38 persons \hspace*{0.2in} \\ \hline
0.0001 & 119 persons \hspace*{0.2in} \\ \hline
$10^{-5}$ & 373 persons \hspace*{0.2in} \\ \hline
$10^{-6}$ & 1,177 persons \hspace*{0.2in} \\ \hline
$10^{-9}$ & 37,229 persons \hspace*{0.2in} \\ \hline
$10^{-12}$ & 1.2 million persons \hspace*{0.2in} \\ \hline
$10^{-15}$ & 37 million persons \hspace*{0.2in} \\ \hline
$10^{-18}$ & 1.2 billion persons \hspace*{0.2in} \\ \hline
\end{tabular}
\end{table}

\section{General Solution for Population Bounds}

The number of pairings possible among $N$ persons is $N(N-1)/2$
because each person can be paired with $N-1$ others, but half of
these are redundant (e.g.\ Alice and Bob, then also Bob and Alice);
hence the halving.  If a biometric technology is operating at some
verification False Match Rate $\mathrm{FMR}$, then the probability of a
given pairing {\em not} resulting in a False Match is $(1-\mathrm{FMR})$,
and the probability that {\em none} of the possible pairings do so is
$(1-\mathrm{FMR})^{N(N-1)/2}$.  For what value of $N$ does this
expression become $< 0.5$, and therefore a biometric collision
becomes likelier than not?

We will invoke a property of the base $e$ ``natural logarithm"
function $\log_{e=2.718...}( \ )$, commonly denoted $\ln( \ )$.
We seek:
\begin{eqnarray}
(1-\mathrm{FMR})^{N(N-1)/2} & < & 0.5 \\ 
\ln \left( (1-\mathrm{FMR})^{N(N-1)/2} \right) & < & \ln(0.5) \\
\frac{N(N-1)}{2} \ln(1-\mathrm{FMR}) & <  & - 0.693 
\end{eqnarray}
Now using the power series expansion
\begin{equation}
 \ln(1+x) = x - \frac{x^2}{2} + \frac{x^3}{3} - \frac{x^4}{4} + \cdots \ ,
\end{equation}
we have $\ln(1+x) \approx x$ for small $|x|$, whether $x \ge 0$ or $x < 0$.
Basically this reflects the fact that the logarithm function is linear
near where it crosses 0 at $\log(1)$, and the slope of this line is 1 if
the base of the logarithm is $e$.  Thus for any small $\mathrm{FMR}$
(say $ < 0.01$), which also entails that $N^{2} \gg N$, we have
\begin{eqnarray}
- \frac{N(N-1)}{2} \ \mathrm{FMR} & \lesssim & - 0.693 \\
N^{2} & \gtrsim & 1.386/\mathrm{FMR} \\
N & \gtrsim & \sqrt{1.386/\mathrm{FMR}} 
\end{eqnarray}

This general (but approximated) solution can be confirmed by evaluating (1)
exactly, using for $N$ each of the corresponding $\mathrm{FMR}$ cases tabulated 
in Table~I, insofar as the available tools of calculation can handle the
combinatorial exponents required in (1) when $N$ is large. 

\section{Biometric Entropy to the Rescue}

Entropy measures the complexity and randomness \cite{CovThom} that is present 
in (and between) random variables.  Facial structure has limited capacity
for randomness.  The major facial features have a canonical standard
configuration, usually with bilateral symmetry; the eyes are normally
on opposite sides of the nose.  Much greater randomness is found in
iris patterns, and this is the origin of their legendary resistance to
False Matches.   Although often there do exist strong radial correlations
within an iris, with mutual information as large as 0.3 bits per bit
across radius \cite{DaugDown2019}, and also IrisCode bits at
adjacent or nearby angles but a shared radial coordinate have ``sticky
oscillator" correlations that reduce their entropy as much as 0.5 bits
per bit \cite{Daug2015}, nevertheless the remaining entropy is vast.
Fig.~1 illustrates this graphically in the bit streams that constitute
the IrisCodes of four different eyes.  How IrisCodes are computed has been
revealed previously \cite{Daug2003}.  The two bit values are equiprobable,
so when bits in IrisCodes from two different eyes are compared by
XOR (Exclusive-OR) to detect whether they agree or disagree, these outcomes
again are equiprobable, amounting to the toss of a fair coin.

\begin{figure}[!h]
\centering
\includegraphics[width=3.5in]{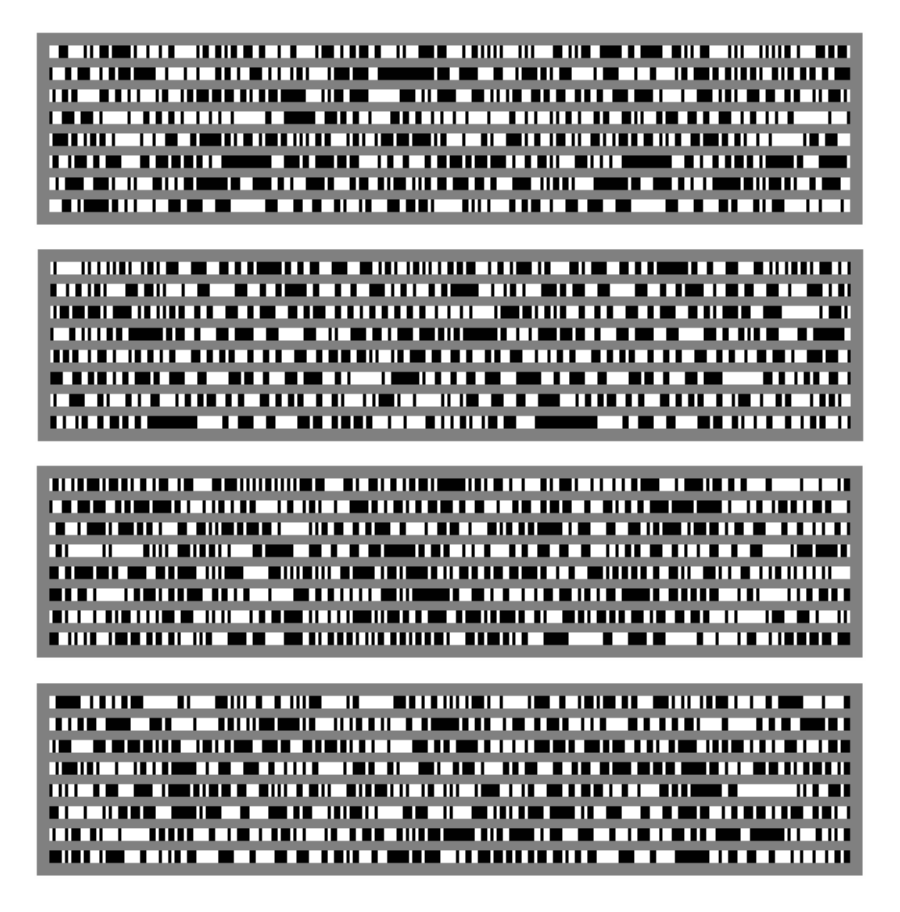}
\caption{Representation of the IrisCodes \cite{Daug2003} produced by four
different eyes.  The eight rows within each can be regarded as eight
concentric rings, each encoding a $[0, 2\pi]$ traversal around the iris.
(Eyelid masking is not shown.)}
\label{fig_iriscodes}
\end{figure}

The non-independence among the bits in a given IrisCode reduces their
collective entropy from what would have been a maximum of 2,048 bits
(if each bit corresponded to an independent ``fair coin toss" Bernoulli
trial) to only about 245 bits.  Modelled as a ``sticky oscillator"
Markov process \cite{Daug2015}, IrisCode bits exhibit a phase coherence
that can persist across several bits.  Despite such losses in entropy,
enough entropy remains that the collision probability between two
IrisCodes from different eyes attenuates by astronomical factors, 
for small reductions in the tolerated fraction of disagreeing bits.

\section{Discussion} 

A good way to understand this effect intuitively is to consider
tossing a fair coin in runs of 245 tosses, tallying each run's fraction
of heads.  The total number of possible outcome sequences is $2^{245}$ 
and each of these has the same probability, namely $p_{i} = 2^{-245}$ 
(including, say, the ``all heads" sequence).  The entropy \cite{CovThom}
contained in these possible sequences is:
\begin{eqnarray}
H & = & -\sum_{i} p_{i} \log_{2} (p_{i})  \\
  & = & -\sum_{i=1}^{2^{245}} 2^{-245} \log_{2}(2^{-245}) = 245 \ \ \mathrm{bits.} 
\end{eqnarray}
The vast majority of these sequences will
have a nearly equal mix of heads and tails.  The fraction
of possible sequences that have (say) fewer than 30\% heads is
less than one-billionth of the total.  This combinatorial property
when large entropy (245~bits) exists in a random variable
is ultimately the reason why, for iris recognition, a match between
two IrisCodes can be accepted even when (say) 30\% of their bits disagree
due to problematic image acquisition.  Despite such a lenient criterion
being so tolerant of noisy bits, the probability that such an accepted
match would actually be a False Match is, indeed, less than 1 in a billion.

The huge exponents appearing in (9) (note that $2^{245} \approx 10^{74}$)
are key to understanding why sufficient entropy is the basis for
biometric collision avoidance even at a planetary scale.  A detailed
tabulation of the relevant probability distributions, both densities
and their cumulatives \cite{Daug2003}, with and without selecting for best 
matches after multiple image rotations to compensate for unknown head and
camera tilt, is provided at \cite{Table245} as a function of Hamming
distance HD (fraction of bits that disagree in IrisCodes from two
different eyes).  This probability table enables us to predict
how tolerant we can be of poor image acquisition (how large a
fraction HD of disagreeing bits we can tolerate and still declare
a match), without resulting in False Matches.  The table \cite{Table245}
shows for acceptance criteria HD the resulting False Match
probability, and its $\log_{10}$ (last two columns).

Table II extracts coarser HD increments of 0.01 from \cite{Table245}
(first column), showing the corresponding FMR predictions
(second column).  By 2003 image databases were only large enough
to perform about 10 million iris cross-comparisons \cite{Daug2003}
but distribution parameters could be estimated, implying 249 bits
of entropy (slightly more than 245), predicting FMR performance
very similar to what is shown in Table~II.  \ No False Matches were
observed below roughly the HD = 0.33 criterion, for the small databases
available.  The predicted FMR values were generally dismissed with incredulity
\cite{Phillips}, because such
FMR performance was unknown in other biometrics.  But subsequently, 
other NIST researchers did actually perform billions \cite{IREX-I}
and then more than a trillion iris comparisons \cite{IREX-III},
obtaining FMR values in good agreement with those predictions,
as reported in column 3.

\begin{table}
\renewcommand{\arraystretch}{1.3}
\caption{False Match Rates Predicted in \cite{Table245}, and as Measured by NIST~\cite{IREX-I}
with 1.16~Billion Iris Comparisons, and \cite{IREX-III} with 1.2~Trillion
Iris Comparisons}
\centering
\begin{tabular}{|c|r|r|}    \hline
{\it HD criterion}  & {\it FMR predicted in} \cite{Table245} & {\it NIST}
\cite{IREX-I}\cite{IREX-III} {\it measured FMR} \\ \hline \hline
0.36 & 1 in 24,000 \hspace*{0.1in} & 1 in 25,000 \hspace*{0.3in} \\ \hline
0.35 & 1 in 110,000 \hspace*{0.1in} & 1 in 71,000 \hspace*{0.3in} \\ \hline
0.34 & 1 in 556,000 \hspace*{0.1in} & 1 in 476,000 \hspace*{0.3in} \\ \hline
0.33 & 1 in 3.1 million \hspace*{0.1in} & 1 in 3.4 million \hspace*{0.3in} \\ \hline
0.32 & 1 in 20 million \hspace*{0.1in} & 1 in 24 million \hspace*{0.3in} \\ \hline
0.31 & 1 in 137 million \hspace*{0.1in} & 1 in 165 million \hspace*{0.3in} \\ \hline
0.30 & 1 in 1.1 billion \hspace*{0.1in} & 1 in 2 billion \hspace*{0.3in} \\ \hline
0.29 & 1 in 9 billion \hspace*{0.1in} & (not measured) \hspace*{0.3in} \\ \hline
0.28 & 1 in 92 billion \hspace*{0.1in} & 1 in 40 billion \hspace*{0.3in} \\ \hline
\end{tabular}
\end{table}

An important cause of skepticism about the FMR performance
levels shown in Table~II, before they were eventually confirmed by 
NIST, was the existence of `ground-truth' errors in early biometric
databases that had created illusory identity collisions.  Apart from 
sloppy and na\"{i}ve data collection, (e.g. incentivising paid student
volunteers to change names and thereby enroll multiple times), there is an
inherent risk in estimating FMR by \emph{intra}-dataset cross-comparisons.  
If even just one of $N$ subjects is enrolled under two different
identities, whether deviously or just through an
innocent clerical error, the estimated FMR then cannot be
better than $2/N^{2}$.  The measured threshold calibration of FMR
such as tabulated in Table~II must then approach a floor, corresponding
to this illusory FMR, which cannot be reduced by any reasonable change 
in threshold, and indeed NIST \cite{IREX-I} demonstrated this problem
for (university-sourced) \emph{intra-}dataset comparisons.
 
NIST overcame this problem by performing \emph{inter-}dataset comparisons:
if two disjoint populations, of sizes (say) $N$ and $M$ in geographically
remote places can be biometrically enrolled, then $N \times M$ inter-comparisons
become possible without the contaminating effect of ground-truth errors. 
NIST \cite{IREX-III} acquired enrollment datasets for two populations
``very well separated geographically and occupationally," one having
3.9 million iris images used as the gallery, and the other having 315,000
iris images used as probes to search against this entire gallery,
asserting there was zero likelihood of co-membership.
Thereby NIST performed $N \times M =$ 1.2~trillion IrisCode comparisons,
leading to the FMR results shown in Table~II column 3 (from \cite{IREX-III}
p.\ 61) for various HD threshold criteria.  This close confirmation of 
theory (column 2), manipulating FMR over a larger than million-fold range,
is striking.

\section{Demographic Specific Application}

Iris pattern entropy differs somewhat across ethnic groups \cite{DDAA}.
For example, the anterior layer of the iris in persons of Sub-Saharan African
descent contains a thick blanket of melanocytes \cite{Snell} creating a
coarser texture of crypts and craters, than the finer fibrous details typically
visible in an iris of persons descended from more northern regions.
Fig.~2 illustrates these entropy differences in samples from three demographies: 
West African; Irish-American; and Nordic.   

\begin{figure}[t]
\centering
\includegraphics[scale=0.92]{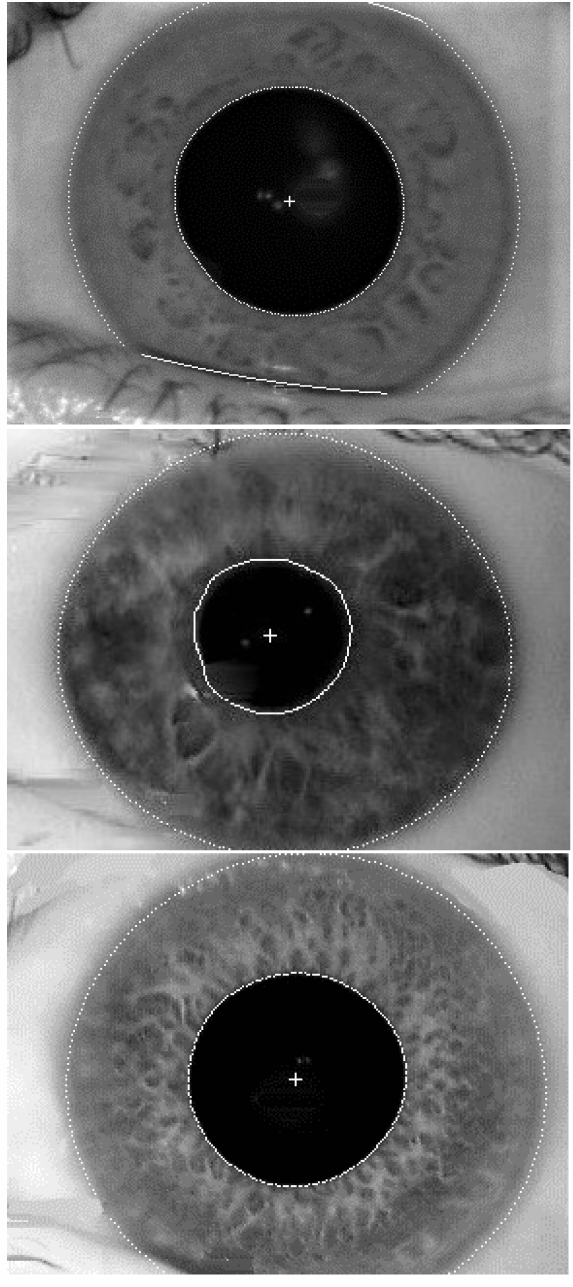}
\caption{Entropy differences in iris patterns from three different 
demographic groups:  West African (top); Irish-American (middle); and Nordic (bottom).}
\end{figure}

Using image databases having particular ethnic demographics, it is possible to
estimate quantitatively their characteristic entropies.  Such calculations are needed
in order to understand how many persons can be enrolled before identity clashes
in ``all-versus-all" cross-comparisons (at a given acceptance operating criterion),
start to become likely.  Fig.~3 illustrates this process for a new West African
database of iris images \cite{DDAA} ``AFHIRIS", plotting the distribution of Hamming distances
(HD, fraction of bits that disagree) between all possible pairings of IrisCodes for different eyes.
The red curve is a plot of the following probability distribution $\mathrm{prob(HD)}$
for the fraction of Heads (HD) in a run of $N$ tosses of a coin whose probability of Heads is $p$\,:
\begin{equation}
\mathrm{prob(HD)} =\frac{N!}{m!(N-m)!}\hspace*{0.08in}p^{m}(1-p)^{(N-m)}
\end{equation}
where in this case $N=228,$ $p=0.5,$ and $\mathrm{HD} = m/N$ is the outcome fraction of $N$ Bernoulli
trials (e.g. observing $m$ Heads within a run of $N$ coin tosses).  Measuring the std dev $\sigma$ for an
empirical distribution of HD scores from independent pairings tells us the equivalent number of
tosses of a coin (having probability $p$ of Heads), namely $N = p(1-p)/\sigma^{2}$.   The empirical
distribution has $\sigma = 0.0331$, with $p \approx 0.5$ (mean~HD) so each toss adds 1 bit of entropy.
Therefore we estimate AFHIRIS biometric entropy as $N = 228$ bits.  The fit in Fig.~3 between the
empirical distribution data and the theoretical probability density curve (10) seems excellent.

\begin{figure}[h]
\centering
\includegraphics[scale=0.42]{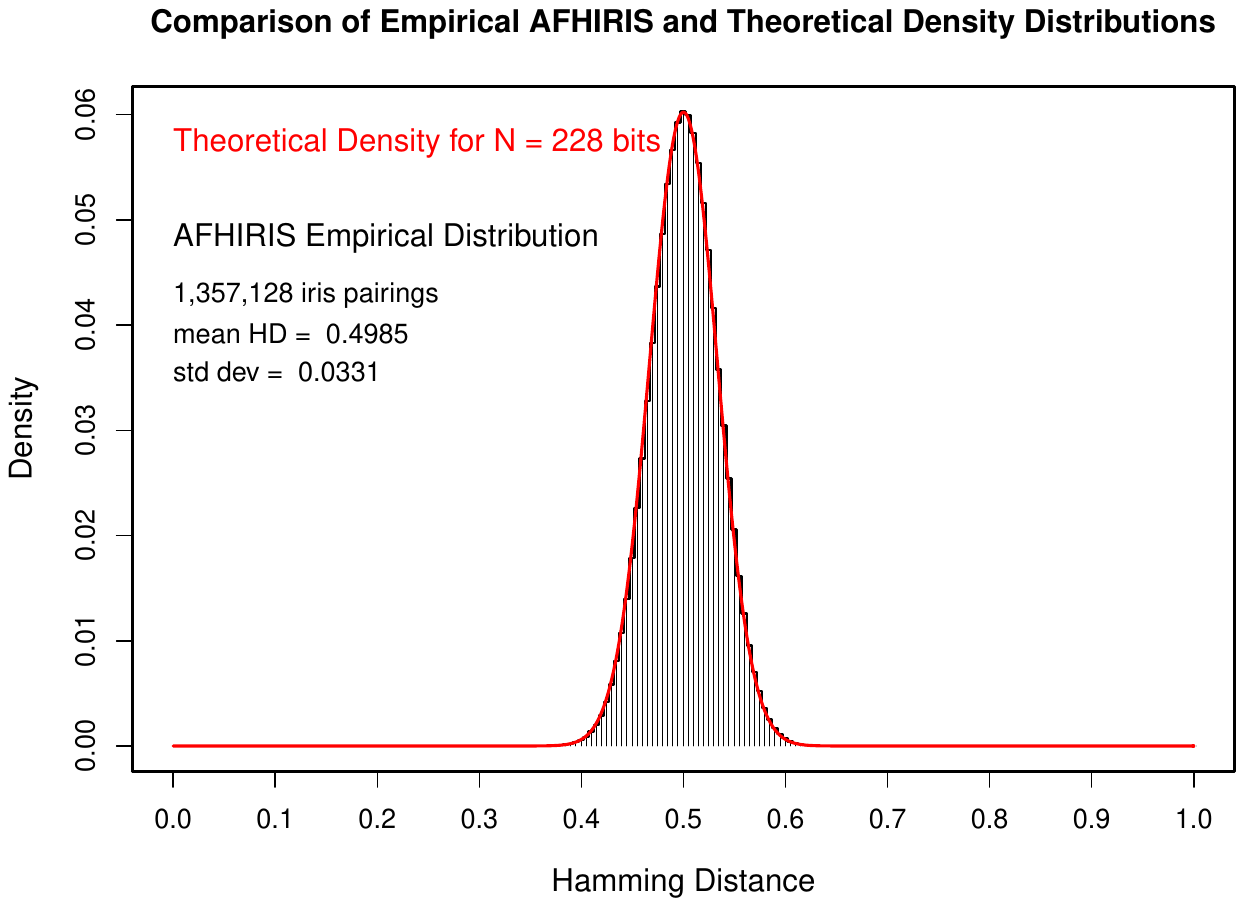}
\caption{Empirical histogram of ``all-versus-all" cross-comparison Hamming distance scores
observed in the West African iris image database AFHIRIS, superimposed with the theoretical bionomial
probability density distribution (red~curve) which plots (10) using parameters $p=0.5$ and $N=228$.}
\end{figure}

As was visible in Fig.~2 and investigated in \cite{DDAA}, biometric entropy in iris patterns varies
among ethnic groups.  The range observed spans from
about 225 bits to 265 bits.  Those values impact the False Match Rates for any given operating point
(with higher entropy reducing the FMR), and therefore they also affect how large a population of 
persons can be enrolled without identity collisions in all-versus-all cross-comparisons.  Such a
concept is sometimes called biometric ``capacity" \cite{Gong} for a given modality and operating
point.  We can now apply the framework that was introduced at the beginning of this paper, the
``biometric birthday problem," to calculate iris capacity across this observed range of entropies.
For any given estimate of biometric entropy, the FMR at a given operating point can be calculated
as described in \cite{Daug2003} and tabulated in \cite{Table245} (for the case of $N = 245$ bits of
entropy).  Using (7) we arrive at the numbers of persons who can be enrolled
while identity collision still remains unlikely.  These numbers are presented in Table III for
two different HD operating thresholds and five estimates of entropy, always assuming single eye
enrollment, to illustrate the combined effects of these variables. 

\begin{table}[h]
\renewcommand{\arraystretch}{1.3}
\caption{Numbers of Persons Enrollable, \ With All-Versus-All Iris
Cross-Comparisons Unlikely To Have Any Identity Collisions, For Two
Operating Points.  \ Single Eye Enrollment Presumed.}
\centering
\begin{tabular}{|c|c|c|}    \hline
{\it Encoded Iris Entropy}  & {\it HD$_{\mathrm{threshold}}$ = 0.28 } & {\it HD$_{\mathrm{threshold}}$ = 0.24 } \\ \hline \hline
225 \ bits & 134,000 \ persons & 16 million \ persons \\ \hline
235 \ bits & 222,000 \ persons & 32 million \ persons \\ \hline
245 \ bits & 370,000 \ persons & 66 million \ persons \\ \hline
255 \ bits & 615,000 \ persons & 136 million \ persons \\ \hline
265 \ bits & 1.02 million \ persons & 278 million \ persons \\ \hline
\end{tabular}
\end{table}

A way to estimate the scalability of face recognition systems was proposed by
\cite{Gong}.  They defined ``face capacity" in terms of packing bounds:  the ratio of the total
volume in a representation space, to the volume that is required to represent individual 
faces in it (as separate spheres or ellipsoids).  This yields an extreme upper bound
estimate of capacity, because there is no way to ensure that the spheres or ellipsoids for
different faces do not overlap.  Such collisions or overlaps certainly occur for identical
twins, and even for unrelated persons who are facial \emph{doppelg\"{a}ngers} (as illustrated
in this collage \cite{doppels} of examples.)  Recent tests by NIST \cite{Grother-face} show
that current face recognition algorithms fail completely to distinguish between identical
twins.   About 1\% of persons have an identical twin, so in any sufficiently broad
sample, face representations must suffer identity clashes for at least those 1\%.
By contrast, it is well-known that the IrisCode produces as much distance between the
encoded iris patterns of identical twins (or indeed between the two eyes of any given
person) as between unrelated eyes \cite{Daug2003}.
 
\section{Conclusion}

Iris recognition is perhaps unique among biometrics in having clear
mathematical foundations, enabling strong predictions about IrisCode
collision likelihood as a function of the decision threshold.
As shown in Table~II, for decision criteria in which no more than
about 31\% of the IrisCode bits are allowed to disagree when declaring
a match (which is a very noise-tolerant criterion), the predicted FMR
attenuates by almost a factor of 10 for each additional 1\% reduction 
in the tolerated amount of bit disagreement.  This extraordinary fact
seems not to be generally understood or appreciated; but it is a direct
result of using high-entropy random variables in biometric codes.
A critical lesson emerging here is the same as a lesson from
cryptography: the great power of randomness, if you can get enough of it.

As confirmed independently by NIST 
in \cite{IREX-III}, the slope of the IrisCode Decision Error Trade-off
curves is so flat that the FMR can be lowered by a factor of 10,000
to 100,000 while not even doubling the False non-Match rate (FnMR).
A consequence of this relationship is that only small costs in
increased FnMR need be paid, by lowering HD threshold, in order to
increase greatly the size of a biometrically enrolled population without
suffering collisions.  Thus for IrisCodes from any two different eyes,
the probability of HD $\le$ 0.29 is about $10^{-10}$.  If we also exploit
the fact that a person's two eyes generate IrisCodes that are almost
completely independent, specifying 0.29 as a match criterion \emph{binocularly}
would yield a fusion FMR of about $10^{-20}$.  Equation (7) shows us
that this is how the planetary human population can survive the
``biometric birthday problem":  it is unlikely that even a single
pairing among 12 billion persons (despite the vast numbers of 
possible pairings) would disagree in $\le$ 29\% of their IrisCode bits
for both pairs of eyes.  Thus speaks biometric entropy.

\bibliographystyle{IEEEtran}

\begin{thebibliography}{12}

\bibitem{Ira}
I.~Kemelmacher-Shlizerman, S.M.~Seitz, D.~Miller, and E.~Brossard, ``The MegaFace
Benchmark:  1 million faces for recognition at scale," \emph{Int'l.\ Conf.\ 
Comp.\ Vision \& Patt.\ Recog.}, pp.\ 4873--4882, 2016.

\bibitem{Grother-face}
P.\ Grother, M.\ Ngan, and K.\ Hanaoka, ``Ongoing Face Recognition Vendor Test (FRVT),
Part 2:  Identification", NISTIR 8238, National Institute of Standards and
Technology (Bethesda), 2018.

\bibitem{PNAS2018}
P.J.~Phillips, A.~Yates, Y.~Hu, C.~Hahn, E.~Noyes, K.~Jackson, J.~Cavazos,
G.~Jeckeln, R.~Ranjan, S.~Sankaranarayanan, J.~Chen, C.~Castillo, R.~Chellappa,
D.~White, and A.J.~O`Toole, ``Face recognition accuracy of forensic examiners,
superrecognizers, and face recognition algorithms," \emph{Proc. Nat'l. Acad. Sci.},
vol.\ 115 (24), pp.\ 6171--6176, 2018.

\bibitem{doppels}
\url{http://www.CL.cam.ac.uk/users/jgd1000/Doppelganger-photos.pdf}

\bibitem{Aiyar}
S.~Aiyar, \emph{AADHAAR: A Biometric History of India's 12-Digit Revolution}.
New~Delhi:  Westland Publications, 2017.

\bibitem{CovThom}
T.~Cover and J.~Thomas,  \emph{Elements of Information Theory}, 2nd~ed.  New~York:
Wiley-Interscience, 2006.

\bibitem{Daug2015}
J.~Daugman, ``Information Theory and the IrisCode,"  \emph{IEEE Trans.\ Info.\ Foren.\ Sec},
vol.\ 11 (2), pp.\ 400--409, 2015.  

\bibitem{DaugDown2019}
J.~Daugman and C.~Downing, ``Radial correlations in iris patterns, and mutual
information within IrisCodes", {\em IET Biometrics}, vol.\ 8 (3), pp.\ 185--189, 2019.

\bibitem{Daug2003}
J.~Daugman, ``The importance of being random:  statistical principles of iris recognition,"
\emph{Pattern Recognition}, vol.\ 36, pp.\ 279--291, 2003.

\bibitem{Table245}
\url{http://www.CL.cam.ac.uk/users/jgd1000/IrisCumulatives.pdf} 

\bibitem{Phillips}
P.J.\ Phillips, \emph{inter alia}, comments at conferences and in publications, 
to UK Government Senior Biometrics Advisors, and to licensees
of these iris recognition algorithms.

\bibitem{IREX-I}
P.~Grother, E.~Tabassi, G.W.~Quinn, and W.~Salamon, ``IREX-I: Performance of Iris Recognition
Algorithms on Standard Images."   \emph{NIST Interagency Report 7629}, 2009.
Data is taken from Fig.~14 (page 46) and Table~7 (page 48) for 1.16 Billion 
inter-dataset iris comparisons.

\bibitem{IREX-III}
P.~Grother, G.W.~Quinn, J.R.~Matey, M.~Ngan, W.~Salamon, G.~Fiumara, and C.~Watson,
``IREX-III:  Performance of Iris Identification Algorithms."   \emph{NIST Interagency Report 7836},
April 6, 2012.

\bibitem{DDAA}
J.~Daugman, C.~Downing, O.N.~Akande, O.C.~Abikoye,
``Ethnicity and biometric uniqueness:  \ iris pattern individuality in a West African database,"
\emph{IEEE Trans.\ Biometrics, Behavior, and Identity Science} (submitted January 2023).

\bibitem{Snell}
R.~Snell, M.~Lemp, \emph{Clinical Anatomy of the Eye} (2nd edition).  London:  Blackwell Science, 1998.

\bibitem{Gong}
S.~Gong, V.N.~Boddeti, and A.K.~Jain, ``On the capacity of face
representation," arXiv:1709.10433, April 2019.

\end{thebibliography}

\begin{IEEEbiography}[{\includegraphics[width=1in,height=1.25in,clip,
keepaspectratio]{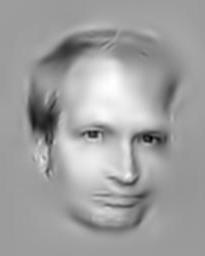}}]{John Daugman}
received his degrees at Harvard University and then
taught at Harvard before coming to Cambridge University, where he is
Professor of Computer Vision and Pattern Recognition.  He has held
the Johann Bernoulli Chair of Mathematics and Informatics at the
University of Groningen, and the Toshiba Endowed Chair at the Tokyo
Institute of Technology.  His areas of research and teaching at
Cambridge include computer vision, information theory, neural computing,
and statistical pattern recognition.  Awards for his work in science 
and technology include the Information Technology Award and Medal of
the British Computer Society, the ``Time 100" Innovators Award, and the
OBE, Order of the British Empire.  He has been elected a Fellow of:    
the Royal Academy of Engineering; the US National Academy of Inventors;
the Institute of Mathematics and its Applications; the British Computer
Society; and he has been inducted into the US National Inventors Hall of
Fame.  He is the founder and benefactor of the Cambridge Chrysalis Trust.
Here he is represented by a sparse sum of 2D Gabor wavelets in six orientations
and five frequencies.
\end{IEEEbiography}

\end{document}